\documentstyle[12pt,graphicx,caption2]{article}

\begin{document}

\title{Models of $n\bar{n}$ transition in medium}
\author{V.I.Nazaruk\\
Institute for Nuclear Research of RAS, 60th October\\
Anniversary Prospect 7a, 117312 Moscow, Russia.}

\date{}
\maketitle
\bigskip

\begin{abstract}
The present state of the $n\bar{n}$ transition problem is briefly outlined. The models based on the diagram technique for direct reactions, potential description of $\bar{n}$-medium interaction and field-theoretical approach are considered. It is shown that for the $n\bar{n}$ transition in medium field-theoretical approach should be used. The lower limit on the free-space $n\bar{n}$ oscillation time $\tau _{{\rm min}}$ is found to be: $10^{16}\; {\rm yr}>\tau _{{\rm min}}>1.2\cdot 10^{9}\; {\rm s}$.
\end{abstract}

\vspace{5mm}

\vspace{5mm}

\vspace{1cm}

*E-mail: nazaruk@inr.ru

\newpage
\setcounter{equation}{0}
\section{Introduction}
At present several models of $n\bar{n}$ transitions in medium are treated. The part of them gives radically different results. This is because the process under study is extremely sensitive to the details of the model and so we focus on the physics of the problem.

We briefly outline the present state of the $n\bar{n}$ transition problem. In the standard calculations of $ab$ oscillations in the medium [1-3] the interaction of particles $a$ and $b$ with the matter is described by the potentials $U_{a,b}$ (potential model). ${\rm Im}U_b$ is responsible for loss of $b$-particle intensity. In particular, this model is used for the $n\bar{n}$ transitions in a medium [4-11] followed by 
annihilation:
\begin{equation}
n\rightarrow \bar{n}\rightarrow M,
\end{equation}
here $M$ are the annihilation mesons.

In [9,10] it was shown that one-particle (potential) model mentioned above does not describe the
process (1) and thus total neutron-antineutron transition probability: the process (1) probability is
$W\sim \Gamma $ (see Eq. (14)), whereas the potential model gives $W\sim 1/\Gamma $ [4-11] ($\Gamma $ is the 
annihilation width of $\bar{n}$ in the medium). In the potential model the effect of final state absorption 
(annihilation) acts in the opposite (wrong) direction, which tends to the additional suppression of the $n\bar{n}$ 
transition. Since the annihilation is the main effect which defines the speed of process (1), the potential model should be rejected. This is because the unitarity condition is used for the essentially non-unitary $S$-matrix [9,10]. The interaction Hamiltonian contains the antineutron optical potential $U_{\bar{n}}$ and ${\rm Im}U_{\bar{n}}$ plays a crucial role. The $S$-matrix should be {\em unitary}.

More formally, the expression for the total process width is obtained by means of optical theorem. The basic equation $\sum_{f\neq i}\mid T_{fi}\mid ^2\approx 2ImT_{ii}$, $S=1+iT$ follows from the unitarity condition $(SS^+)_{fi}=\delta _{fi}$. However in the potential model the $S$-matrix is {\em essentially} non-unitary $(SS^+)_{fi}=\delta _{fi}+\alpha _{fi}$, $\alpha _{fi}\neq 0$, resulting in $\sum_{f\neq i}\mid T_{fi}\mid ^2\approx 2ImT_{ii}+\alpha _{ii}\neq 2ImT_{ii}$ because $2ImT_{ii}$ is extremely small: $2ImT_{ii}<10^{-31}$ [9,10]. The above-given basic equation is inapplicable in this case. The potential model describes only the channel with $\bar{n}$ in the final state [10] when unitarity condition is not used.

For the oscillations in the external field [12,13] the Hamiltonian is hermitian and so there is no similar problem. The above-given remark holds only for the processes (1) and total neutron-antineutron transition probability calculated by means of  non-hermitian Hamiltonian (potential model).

The potential model was developed in 1980-1981. In more recent papers the verious details of the model have been refined, in particular the parameters of optical potential. We don't dwell on these papers since the heart of the problem is in the non-Hermiticity of the optical potential. For similar reason for the model based on the diagram technique (see below) we consider papers [14,15] only. 
 
In [14] we have proposed the model based on the diagram technique for direct reactions (see Fig. 1). This model (later on referred to as the model 1) does not contain the non-hermitian operators. For deuteron this calculation was repeated in [15]. However, in [16] it was noted that this model is unsuitable for the problem under study. In Sect 2 this problem is considered in detail. The model shown in Fig. 2 (model 2) has been proposed [11,16]. 

In the model 2 the antineutron propagator can be bare or dressed. The corresponding results differ radically for the following reason.  A distinguishing feature of the problem under study is the zero momentum transfer in the $n\bar{n}$ transition vertex. Because of this the $S$-matrix amplitudes corresponding to the processes shown in Figs. 2a and 2b with bare propagator contain infrared divergence (see Eq. (9)). The small value of the antineutron self-energy $\Sigma $ removes the singularity from the process amplitude which changes the result radically. This circumstance demonstrates the result sensitivity to the details of the model. As a consequence the lower limit on the free-space $n\bar{n}$ oscillation time $\tau _{{\rm min}}$ given in this paper is in the wide range: $10^{16}\; {\rm yr}>\tau _{{\rm min}}>1.2\cdot 10^{9}\; {\rm s}$. In addition, certain of the authors consider that the model 1 (and even potential model) is reasonable. From the preceding, it is seen that the problem invites further investigation. 

In this paper we compare the model based on diagram technique for direct reactions (section 2) with the the model based on the field-theoretical approach (section 3) and show that the former model is unsuitable for the problem under study. The model 2 and in the first place the value of antineutron self-energy invites further investigation. By means of  field-theoretical approach the range of permissible values of $\tau _{{\rm min}}$ is obtained (section 3). Section 4 contains the conclusion.

\section{Model 1}
Consider now the model 1 [14]. The Hamiltonian of $n\bar{n}$ transition is [4]
\begin{equation}
{\cal H}_{n\bar{n}}=\epsilon \bar{\Psi }_{\bar{n}}\Psi _n+H.c.
\end{equation}
Here $\epsilon $ is a small parameter with $\epsilon =1/\tau _{n\bar{n}}$, where $\tau 
_{n\bar{n}}$ is the free-space $n\bar{n}$ oscillation time.

\begin{figure}[h]
  {\includegraphics[height=.25\textheight]{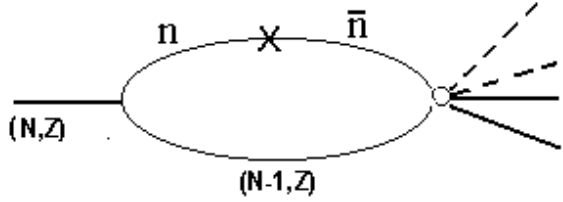}}
  \caption{Model 1 for the $n\bar{n}$ transition in the nuclei followed by annihilation.}
\end{figure}

Denote: $A=(N,Z)$, $B=(N-1,Z)$ are the initial and intermediate nuclei, $M$ is the
amplitude of virtual decay $A\rightarrow n+(A-1)$, $M_a^{(n)}$ is the amplitude of 
$\bar{n}B$ annihilation in $(n)$ mesons, $E_n$ is the pole neutron binding energy; $m$, 
$m_A$, $m_B$ are the masses of the nucleon and nuclei $A$ and $B$, respectively. The 
amplitude is given by 
\begin{equation}
M^{(n)}=-\frac{im^2m_B}{2\pi ^4}\epsilon 
\int d{\bf q}dE \frac{M(q)M_a^{(n)}(m_A)}{({\bf q}^2-2mE-i0)^2[{\bf q}^2+2m_B
(E+E_n)-i0)]}.
\end{equation}
For deutron the process probability $W_1(t)$ was found to be
\begin{equation}
W_1(t)=\frac{\epsilon ^2}{6E_n^2}\Gamma _{\bar{n}p}t,
\end{equation}
where $\Gamma _{\bar{n}p}$ is the $\bar{n}p$ annihilation width. For carbon and oxygen
$W_1\sim \Gamma _{\bar{n}B}t/E_n^2$ as well ($\Gamma _{\bar{n}B}$ is the width of $\bar{n}B$ 
annihilation). The lower limit on the free-space $n\bar{n}$ oscillation time which follows from stability of oxygen is [14] 
\begin{equation}
\tau ^1_{{\rm min}}({\rm O})=1.1\cdot 10^{8}\; {\rm s}.
\end{equation}
The limit obtained from stability of Fe is $\tau ^1_{{\rm min}}({\rm Fe})=(8\div 11)\cdot 10^{8}\; {\rm s}$ [15].

We list the main drawbacks of the model which are essential for the problem under study:

1) The model does not reproduce the $n\bar{n}$ transitions in the medium and vacuum. If the 
neutron binding energy goes to zero, Eq. (4) diverges (see also Eqs. (15) and (17) of 
Ref. [15]). 

2) Contrary to the model 2 (see next section), the amplitude (3) cannot be obtained from the Hamiltonian because in the interaction Hamiltonian {\em there is no term} which induces the virtual decay $(N,Z)\rightarrow n+(N-1,Z)$. The neutron of the nucleus is in the bound state and so it should be described by the wave function and not the propagator.

3) The model does not contain the infrared singularity for any process including the 
$n\bar{n}$ transition, whereas it exists for the processes in the medium and vacuum (see
[11,16] and Sect. 3 of this paper). This brings up the question: Why? The answer is that 
for the propagator the infrared singularity cannot be in principle since the particle is 
virtual: $p_0^2\neq m^2+{\bf p}^2$. Due to this the model is infrared-free.

4) Since the model is formulated in the momentum representation, it does not describe the 
coordinate-dependence, in particular the loss of particle intensity due to absorption.
Also there is a no the dependence on nuclear density! The model is crude and has very restricted range of applicability. 

We consider the points 2) and 3). The $n\bar{n}$ transition takes place in the propagator. As the result the model is infrared-free. For the processes with zero (or very small)
momentum transfer this fact is crucial since it changes the functional structure of the amplitude.

On the other hand, the neutron propagator arises owing to the vertex of virtual decay $A\rightarrow n+(A-1)$. However, as pointed out above, in the interaction Hamiltonian {\em there is no term} which induces the virtual decay $A\rightarrow n+(A-1)$. This vertex is the artificial element of the model. It was introduced in order for the neutron (pole particle) state to be separated.

We assert that for the problem under study this scheme is incorrect. The neutron state is
described wrongly. The diagram technique for direct reactions has been developed and adapted to the direct type reactions. The term "diagram technique for direct reactions"
emphasizes this circumstance. The processes with non-zero momentum transfer are considerably less sensitive to the description of pole particle state. The approach is very handy, useful and simple since it is formulated in the momentum representation. This approach was applied by us for the calculation of knock-out reactions and $\bar{p}$-nuclear annihilation [17]. The price of simplicity is that its applicability range is {\em restricted}. At the same time, as is seen from p. 3), the process under study is extremely sensitive to the description of neutron state. The same is true for the value of antineutron self energy (see next section). Besides, the problem is unstable [18].

\section{Model 2}
Model 2 describes both $n\bar{n}$ transition in the medium (Fig. 2a) and nuclei (Fig. 2b).
It corresponds to the standard formulation of the problem: the $|in>$-state is the 
eigenfunctions of unperturbed Hamiltonian. In the case of the process shown in Fig. 2a, 
this is the neutron plane wave:
\begin{equation}
n_p(x)=\Omega ^{-1/2}\exp (-ipx),   
\end{equation}
$p=(\epsilon _n,{\bf p}_n)$, $\epsilon _n={\bf p}_n^2/2m+U_n$, where $U_n$ is the neutron potential. In the case of Fig. 2b, this is the wave function of bound state [19]. For the nucleus in the initial state we take the one-particle shell model. Since the neutron is described by the bound state wave function and not the plane wave, the antineutron Green function differs from that in Fig. 2a. To emphasize this fact, for the
virtual antineutron in Figs. 2b and 2a the different terms are used. 

\begin{figure}[h]
  {\includegraphics[height=.25\textheight]{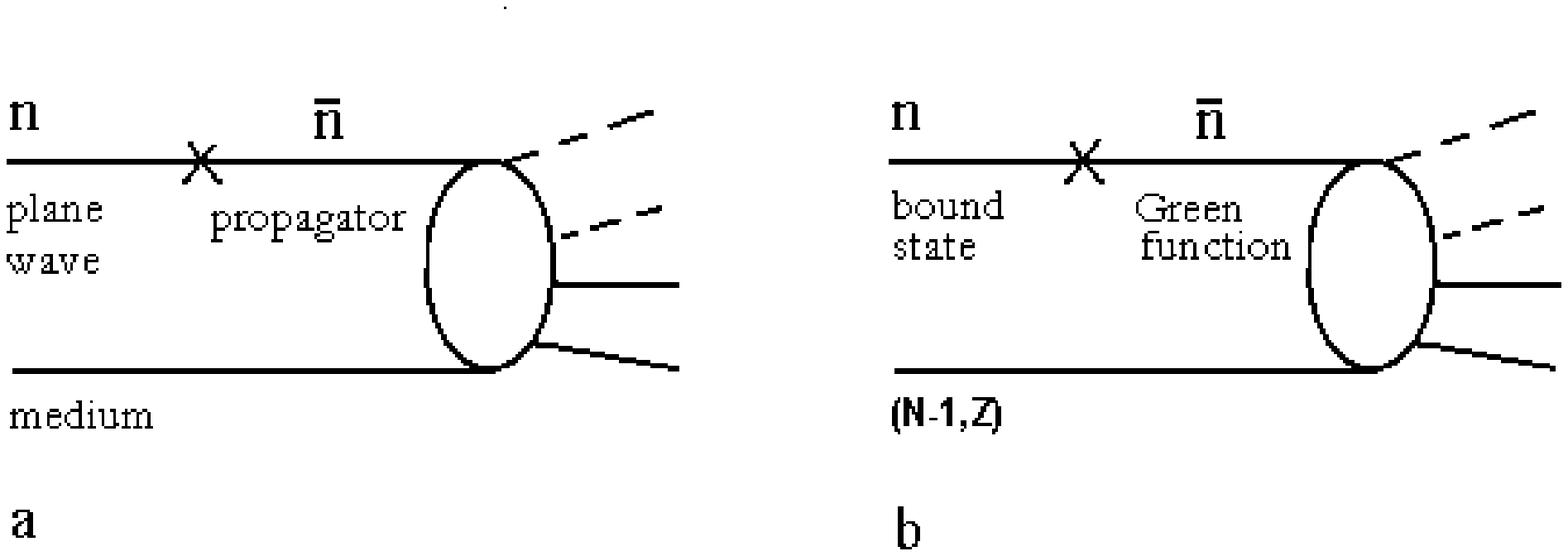}}
  \caption{Model 2 for the $n\bar{n}$ transition in the medium ({\bf a}) and nuclei ({\bf b}) followed by annihilation.}
\end{figure}

The interaction Hamiltonian is given by
\begin{equation}
{\cal H}_I={\cal H}_{n\bar{n}}+{\cal H},
\end{equation}
where ${\cal H}$ is the hermitian Hamiltonian of $\bar{n}$-medium interaction in the case
of Fig. 2a and Hamiltonian of $\bar{n}$-nuclear interaction in the case of Fig. 2b.
In so doing the antineutron propagator can be bare or dressed.

\subsection{Model 2 with bare propagator}
Consider now the diagram 2a. We show the result sensitivity to the details of the model.
If we use the general definition of amplitude of antineutron annihilation in the 
medium $M_{ann}$ which is given by
\begin{eqnarray}
<\!f\!\mid T^{\bar{n}}\mid\!0\bar{n}_p\!>=N(2\pi )^4\delta ^4(p_f-p_i)M_{ann},\nonumber\\
T^{\bar{n}}=T\exp (-i\int dx{\cal H}(x))-1
\end{eqnarray}
($\mid\! 0\bar{n}_p\!>$ is the state of the medium containing the $\bar{n}$ with the 
4-momentum $p$, $N$ includes the normalization factors) then the amplitude corresponding to 
Fig. 2a diverges:
\begin{eqnarray}
M_b=\epsilon G_bM_{ann},\nonumber\\
G_b=\frac{1}{\epsilon _{\bar{n}}-{\bf p}_{\bar{n}}^2/2m-U_n}\sim \frac{1}{0}
\end{eqnarray}
since ${\bf p}_{\bar{n}}={\bf p_n}$, $\epsilon _{\bar{n}}=\epsilon _n$. This is infrared singularity conditioned by zero momentum transfer in the $n\bar{n}$ transition vertex. Since $M_{ann}$ contains all the $\bar{n}$-medium interactions followed by annihilation including antineutron rescattering in the initial state, the antineutron propagator $G_b$ is bare. The index "b" emphasizes this fact. Once the antineutron annihilation amplitude is defined by (8), the expression for the process amplitude (9) {\em rigorously follows} from (7). It is significant that the observable values (antineutron annihilation width $\Gamma $ for example) are expressed through $M_{ann}$.

For solving the problem the field-theoretical approach with finite time interval is used [20]. 
It is infrared free. For the probability of the process (1) we get [11,16]
\begin{equation}
W_b(t)\approx W_f(t)=\epsilon ^2t^2, \quad \Gamma t\gg 1,
\end{equation}
where $W_f$ is the free-space $n\bar{n}$ transition probability. Equation (10) does not contain density-dependence ($\Gamma-$dependence) since it corresponds to the limiting case $\Gamma t\gg 1$ which is realized for the $n\bar{n}$ transition in nuclei.

The lower limit on the free-space $n\bar{n}$ oscillation time is found to be
\begin{equation}
\tau ^b_{{\rm min}}=10^{16}\; {\rm yr}.
\end{equation}
This value is interpreted as the estimation from above.

\subsection{Model 2 with dressed propagator}
In the case considered above the amplitude $M_{ann}$ involves all the $\bar{n}$-medium 
interactions followed by annihilation including the antineutron rescattering in the initial 
state. In principle, the part of this interaction can be included in the antineutron
propagator. Then the antineutron self-energy $\Sigma $ is generated and
\begin{equation}
G_b\rightarrow  G_d=\frac{1}{\epsilon _{\bar{n}}-{\bf p}_{\bar{n}}^2/2m-U_n-\Sigma }=-\frac{1}{\Sigma }.
\end{equation}
The process amplitude $M_d$ is non-singular:
\begin{equation}
M_d=\epsilon G_dM'_{ann}.
\end{equation}
The parameter $\Sigma $ and "amplitude" of antineutron annihilation $M'_{ann}$ are uncertain. The annihilation width $\Gamma $ is expressed through $M_{ann}$ and not $M'_{ann}$. Compared to $M_{ann}$, $M'_{ann}$ is truncated because the portion of the Hamiltonian ${\cal H}$ is included in $G_d$. 

For the estimation we put $M'_{ann}=M_{ann}$. This is an uncontrollable approximation.
The process probability is found to be
\begin{eqnarray} 
W_d(t)\approx \Gamma _dt,\nonumber\\
\Gamma _d\approx \frac{\epsilon ^2}{\Sigma ^2}\Gamma .
\end{eqnarray}
The result differs from (10) fundamentally. Let $\tau ^d_{{\rm min}}$ be the lower limit on the free-space $n\bar{n}$ oscillation time obtained by means of Eq. (14); $T_{n\bar{n}}$ is the oscillation time of neutron bound in a nucleus. Using condition $W_d(T_{n\bar{n}})<1$, one obtains
the relationships between $\tau ^d_{{\rm min}}$ and $T_{n\bar{n}}$:
\begin{equation}
\tau ^d_{{\rm min}}=\frac{1}{\Sigma }\sqrt{\Gamma T_{n\bar{n}}}.
\end{equation}
We use the experimental bound on the neutron lifetime in oxygen $T_{n\bar{n}}>1.77\cdot 10^{32}$ yr obtained by Super-Kamiokande collaboration [8]. For estimation we take $\Gamma =100$ MeV and $\Sigma \approx {\rm Re}U_{\bar{n}}-U_n\approx 10$ MeV ($U_n$ and $U_{\bar{n}}$ are the potentials of $n$ and $\bar{n}$, respectively). Then  
\begin{equation}
\tau ^d_{{\rm min}}=1.2\cdot 10^{9}\; {\rm s},
\end{equation}
which exceeds the lower limit given by the Grenoble reactor experiment [21] by a factor of 14 and the restriction given by potential model [8] by a factor of five. Limit (16) is comparable to the restriction $\tau ^1_{{\rm min}}({\rm Fe})=(8\div 11)\cdot 10^{8}\; {\rm s}$ given in [15], however this coincidence is accidental. The model 1 should be rejected. If $\Sigma \rightarrow 0$, $W_d$ rises quadratically. So $\tau ^d_{{\rm min}}$ can be considered as the estimation from below.
 
In the models with bare and dressed propagators the interaction Hamiltonians ${\cal H}_I$ and unperturbed Hamiltonians are the same. The sole physical distinction between above-mentioned models is the zero antineutron self-energy in the model with bare propagator; or, similarly, the definition of antineutron annihilation amplitude ($M_{ann}$ or $M'_{ann}$). However, it leads to the fundamentally different results. The problem is extremely sensitive to the value of antineutron self-energy $\Sigma $ (see (9), (10) and (13), (14)) as well as the description of initial neutron state (propagator in the model 1 or wave function in the model 2) and the value of momentum transferred in the $n\bar{n}$ transition vertex [18]. This is because the amplitude (9) is in the {\em peculiar point}. 

On the one hand, the value $\Sigma =0$, i. e. the model with bare propagator, seems quite realistic for pure physical reasons [11,18], and on the other hand the result is extremely sensitive to the details of the model. Besides, the calculation corresponding to the model 2 with bare
propagator contains too many new elements and we view the result (10) with certain caution. Due to this we adduce the wide range of permissible values of $\tau _{{\rm min}}$:
\begin{equation}
\tau ^b_{{\rm min}}>\tau _{{\rm min}}>\tau ^d_{{\rm min}}.
\end{equation}
This is a problem of great nicety. Further investigations are desirable. 

For the neutrons in bound state (see Fig. 2b) the results are the same as for nuclear matter [19].

\section{Summary and conclusion}
Since the operator (2) acts on the neutron, in the model 1 the vertex of virtual decay $A\rightarrow n+(A-1)$ is introduced because one should separate out the neutron state. This scheme is artificial because in the interaction Hamiltonian {\em there is no term} which induces the virtual decay $A\rightarrow n+(A-1)$. The neutron of the nucleus is in the bound state and so it cannot be described by the propagator.
In more exact terms, this is a crude model which is inapplicable for the problem under study. (This point is discussed more comprehensively in 
[22].) Alternative method is given by the model 2 which does not contain the above-mentioned vertex. The $|in>$-states are the eigenfunctions of unperturbed Hamiltonian what has no need of a commentary.

It should be emphasized that oscillations in the vacuum and gas, in particular oscillations of ultra cold neutrons [6] are considered as well. These processes are described in the framework of the model shown in Fig. 2a [11,18]. They cannot be reproduced by means of model 1 in principle. For the reasons given above the model 1 should be rejected. 
If $\Sigma \rightarrow 0$, $W_d(t)$ diverges quadratically. This circumstance should be clarified; otherwise the model 2 can be rejected. The calculation in the framework of the model with bare propagator gives the finite result, which justifies our approach from a conceptual point of view and consideration of the model with bare propagator at least as the limiting case. In reality this model seems quite realistic in itself [11,18,19].

Finally, taking into account the result sensitivity to the description of neutron state, we argue that the model 1 is inapplicable to the problem under study. The same is true for the potential model (see section 1). The model 2 and in the first place the value of antineutron self-energy invites further investigation. At present one can adduce only wide range of permissible values of $\tau _{{\rm min}}$. If $\Sigma $ changes in the limits $10\; {\rm MeV}>\Sigma >0\; {\rm MeV}$, the lower limit on the free-space $n\bar{n}$ oscillation time is in the range
\begin{equation}
10^{16}\; {\rm yr}>\tau _{{\rm min}}>1.2\cdot 10^{9}\; {\rm s}.
\end{equation}

Recall that for the free-space $ab$ oscillations the $ab$ transition probability is extremely sensitive to the difference of masses $m_a-m_b$
as well. The fact that process amplitude is in the peculiar point is the basic reason why the problem is extremely sensitive to $\Sigma $ and the range of permissible values of $\tau _{{\rm min}}$ is very wide. The values $\tau _{{\rm min}}=1.2\cdot 10^{9}\; {\rm s}$ and $\tau _{{\rm min}}= 10^{16}$ yr are interpreted as the estimations from below (conservative limit) and from above, respectively. The estimation from below exceeds the restriction given by the Grenoble reactor experiment [21] by a factor of 14 and the lower limit given by potential model [8] by a factor of 5. The range of uncertainty of $\tau _{{\rm min}}$ is too wide. Further theoretical and experimental investigations are desirable.

\end{document}